# Creation of Optical Cat and GKP States Using Shaped Free Electrons

Raphael Dahan[†1], Gefen Baranes[†1], Alexey Gorlach[1], Ron Ruimy[1], Nicholas Rivera[2], and Ido Kaminer[*1]

[1] *Technion – Israel Institute of Technology, Haifa 32000, Israel*

[2] *Massachusetts Institute of Technology, Cambridge MA 02139, USA*

[†]*equal contributors*   [*]*kaminer@technion.ac.il*

**Cat states and Gottesman-Kitaev-Preskill (GKP) states play a key role in quantum computation and communication with continuous variables. The creation of such states relies on strong nonlinear light-matter interactions, which are widely available in microwave frequencies as in circuit quantum electrodynamics platforms. However, strong nonlinearities are hard to come by in optical frequencies, severely limiting the use of continuous variable quantum information in the optical range. Here we propose using the strong interactions of free electrons with light as a source for optical cat and GKP states. The strong interactions can be realized by phase-matching of free electrons with photonic structures such as optical waveguides and photonic crystals. Our approach enables the generation of optical GKP states with above 10 dB squeezing and fidelities above 90% at post-selection probability of 10%, even reaching >30% using an initially squeezed vacuum state. Furthermore, the free-electron interaction allows for conditional rotations on the photonic state, enabling to entangle a pair of GKP states into a GKP Bell state. Since electrons can interact resonantly with light across the electromagnetic spectrum, our approach may be used for the generation of cat and GKP states over the entire electromagnetic spectrum, from radio-waves to X-rays.**

# Introduction

The fundamental interaction between an electromagnetic field and free electrons forms the basis of electrodynamics. The radiation emitted by the electrons can span a wide spectral range, extending from radio-waves all the way up to X-rays, as in cyclotron radiation, traveling wave tubes, the Cherenkov effect, X-ray tubes, and free-electron lasers [1]. The classical properties of the radiation, such as intensity and frequency, can be controlled by the classical electron characteristics such as its acceleration along a trajectory. Control of such electron characteristics has been the cornerstone of free-electron radiation technologies, using either static electromagnetic fields as in synchrotron radiation [2-4] or oscillating fields as in Compton scattering [2,5,6].

Importantly, ever since the work of Glauber and other luminaries [7-9], it is widely appreciated that the full nature of light goes far beyond its classical characteristics. There are also quantum degrees of freedom, such as entanglement and the degrees of coherence. In general, realizing full control of the quantum state of light is a long-standing open problem in quantum optics. In particular, substantial effort is invested toward an efficient generation of the desired quantum states that enable fault-tolerant quantum computing with continuous variables (CVs) [10-13]. Prominent examples are the Gottesman-Kitaev-Preskill (GKP) states [10], which are designed to be robust against errors. Recently, such states were demonstrated in the microwave frequency range in platforms of circuit quantum electrodynamics and ion traps [14-16]. More generally, all the states that enable universal quantum computation with CVs are part of the so-called non-Gaussian states [11,17] (states whose Wigner quasi-probability distribution is not a Gaussian function). Unfortunately, the desired non-Gaussian states such as GKP states are yet to be demonstrated in the optical range, due to the generally weak nature of optical nonlinearities [18,19].

Current theoretical proposals for the generation of optical GKP states rely on optical Kerr effects [20,21], cavity QED [22,23], homodyne measurements of cat states [24,25], or measurement-based schemes that require photon-number-resolving detection [19,26,27]. However, all these optical schemes are currently limited to the generation of just a few photons with low fidelities, because they rely on the intrinsically weak optical nonlinearities or on low post-selection probabilities. This is why no optical experiment so far has reached a non-Gaussian state of sufficient photon-number and sufficient squeezing to be usable for CV quantum computation. Despite the difficulties in creating optical GKP states, there is an ongoing intense search for new mechanisms to generate them and unlock their prospects for photonic quantum technologies.

Here we propose to exploit interactions between free electrons and photonic structures to generate GKP states and other non-Gaussian states that can facilitate fault-tolerant quantum computing. Our approach provides control over the quantum state of free-electron radiation by pre-shaping the electron wavefunction before its radiation emission and post-selecting the electron energy afterward. Specifically, we propose energy-comb electrons as a natural basis for controlling the photonic states created via the interaction. A free-electron comb is a superposition of electron energy eigenstates in which the energies form an evenly spaced ladder, analogously to an optical frequency comb of evenly spaced frequencies. We show how cat states and multi-component cat states can be heralded by energy post-selection of comb electrons. Consequent interactions of multiple comb electrons with appropriate post-selections create more complex photonic states such as the GKP state. We find that the post-selection probability to produce a GKP state of 10 dB squeezing is >10%, on par with current leading theoretical proposals for the creation of optical GKP states [19-27]. We further present more advanced schemes that increase the probability to >30% by "seeding" the electron radiation process with a squeezed vacuum state that can be generated using spontaneous parametric processes [28]. Finally, we demonstrate how the interactions with comb electrons can apply gates on the GKP states, for example performing a conditional rotation that entangles two photonic modes into a GKP Bell state – an important step toward the vision of GKP cluster states for fault-tolerant quantum computation.

Earlier papers suggested [29] and demonstrated [30] that post-selection on shaped free electrons alters the properties of their emission. In a different approach, without pre-shaping, free electrons were proposed as single-photon emitters [31] and recently utilized in an experiment to create quantum-correlated electron-photon pairs [32]. Moreover, photon addition or subtraction through free-electron post-selection was proposed for generating Fock states, photon-added coherent states, or photon-subtracted thermal states [33]. Other recent works unveiled the dependence of the second-order coherence of the emitted light on the electron's wavefunction duration and shape [34-37]. A significant research effort over the past decade explored the shaping of the single electron wavefunction in the longitudinal [38-42] and transverse [41-53] directions. Most importantly for our approach, time-energy shaping of a free-electron wavefunction was demonstrated by ultrafast transmission electron microscopy (UTEM) [38,39]. Such temporal shaping has shown in recent years the creation of coherent free-electron attosecond bunches [40] and free-electron energy combs [54,55]. These advances show the feasibility of the concept we suggest here.

## Results

**Creation of quantum state of light using coherent free-electron combs**

The process we propose for creating desired non-Gaussian quantum states consists of three building blocks (Fig. 1): (a) generation of shaped electrons (i.e., comb electrons), (b) efficient free-electron-photon interaction (in the strong coupling regime), and (c) electron energy post-selection. We assume a highly paraxial electron with energy much higher than that of the photon with which it interacts, yet energy uncertainty smaller than the photon energy. This condition is frequently realized in transmission and scanning electron microscopes, as exemplified by different experiments in photon-induced near-field electron microscopy (PINEM) [38-41,51,53-59], and explained theoretically in [60,61].

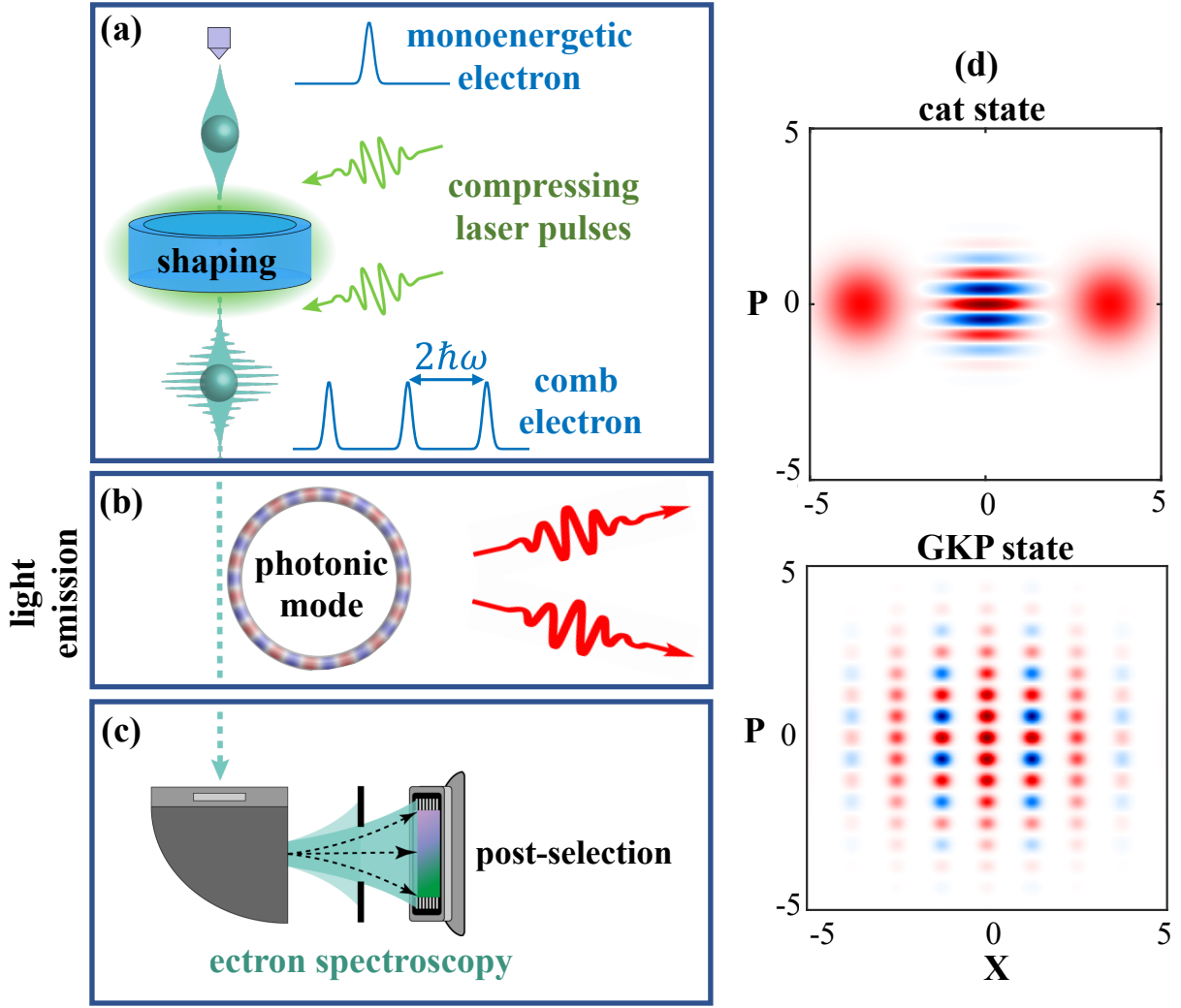

**Fig. 1 | Generating optical cat and GKP states using free electrons.** A scheme of free-electron-based cat and GKP state generation, divided into three main stages: **(a)** Preparation stage: a monoenergetic electron is shaped into a comb with an energy spacing of $N \cdot \hbar\omega$. The case illustrated here shows $N = 2$. **(b)** Light-emission stage: the electron then interacts with a photonic structure, emitting photons into an optical mode. The preferred structures are ones designed to guide light in a waveguide/cavity that support a phase-matched interaction with the electron for a stronger coupling constant (higher $g_Q$). **(c)** Post-selection stage: the electron is measured, heralding the generated photonic state. In the case illustrated here, if the measured energy is even ($k = 0$), an even cat state is created $|\alpha\rangle + |-\alpha\rangle$. If the measured energy is odd ($k = 1$), an odd cat state is created $|\alpha\rangle - |-\alpha\rangle$. **(d)** Examples of Wigner functions for the photonic states that can be generated via our scheme: A cat state can be generated using an electron comb and a post-selection of its energy. A GKP state can be generated using multiple comb electrons with a specific sequence of post-selections as described in Table 1. ($g_Q = 4$ and $\sqrt{\pi/2}$ for the top and bottom panels, respectively).

Using a quantum-optical framework, the interaction between a free electron and an optical mode is captured by the scattering matrix (as first proposed in [62,63] and demonstrated experimentally in [57]):

$$S = \exp[g_Q b a^\dagger - g_Q^* b^\dagger a], \tag{1}$$

where $a, a^\dagger$ are the annihilation and creation operators for the photonic mode; $b, b^\dagger$ (satisfying $bb^\dagger = b^\dagger b = 1$) are operators describing an electron translation in energy, which correspond to the emission or absorption of a single photon. A general electron wavefunction is described as a superposition of monoenergetic states $|n\rangle_e = |E_0 + n\hbar\omega\rangle_e$, each describing an electron shifted by a multiple of the photon energy $n\hbar\omega$. The condition to consider the electron as occupying such a discrete ladder of energy states is that the electron interacts predominantly with a single optical mode of frequency $\omega$. In conventional PINEM experiments that probe stimulated interactions, this condition is ensured by the pump laser linewidth, which creates a narrow-bandwidth excitation. In the spontaneous (non-pumped) case we consider here, the condition to consider the electron as on a discrete ladder is that it predominantly emits into a single optical mode. Using this notation, the electron translation operators satisfy $b^\dagger |n\rangle_e = |n+1\rangle_e$, $b|n\rangle_e = |n-1\rangle_e$. We choose the state $|0\rangle_e$ as the initial electron state before it is shaped into a comb.

The coupling constant $g_Q$ is a dimensionless complex parameter that describes the interaction strength and the phase between the optical mode and the free electron. $g_Q$ is defined using the electric field $E$ of the optical mode, normalized to the amplitude of a single photon, with $v$ being the electron velocity and $r_\perp$ being its transverse location: $g_Q = (q_e/\hbar\omega) \int E_z(r_\perp, z) e^{-i\omega z/v} dz$ [62]. Equivalently, $g_Q$ can also be derived from the Green's function of the optical structure [64]: $|g_Q|^2 = (q_e^2 \mu_0 / \pi\hbar) \int \mathrm{Im} G_{zz}(r_\perp, z; r_\perp, z'; \omega) e^{-i\omega(z-z')/v} dz dz'$.

The free-electron wavefunction can be shaped in the time domain, i.e. undergo a temporal modulation, induced by the interaction as in PINEM [38-41,51,53-59] or the pondermotive interaction [65-67]. In this paper, we consider electrons shaped as energy combs with a periodicity of multiple photon energy $N \cdot \hbar\omega$ and equal phases (reference material (RM) 1). Such an ideal electron comb can be approximated by shaping a monoenergetic electron using multiple frequencies [68] or multiple interaction stages [69]. The next section shows that these combs can be used for heralding different cat states, under certain energy post-selection conditions.

**Creation of N-component cat states**

Let us focus on the creation of multi-component cat states [9] (also known as multi-legged cat states), which can be described as $|\text{cat}_N^k\rangle_{\text{ph}} \propto \sum_{m=0}^{N-1} \exp(-2\pi i k m/N)|\exp(2\pi i m/N)\alpha\rangle_{\text{ph}}$, where $|\alpha\rangle_{\text{ph}}$ describes a coherent state. To create the N-component cat state, we prepare a comb electron $|\text{comb}_N^0\rangle_e \propto \sum_{n=-\infty}^{\infty}|E_0 + n \cdot N \cdot \hbar\omega\rangle_e$ with an energy spacing of $N \cdot \hbar\omega$ (Fig. 2a). We further define the shifted combs to be $|\text{comb}_N^m\rangle_e = b^{\dagger m}|\text{comb}_N^0\rangle_e$, noting that any $|\text{comb}_N^m\rangle_e$ is invariant under $b^N$. Assuming an initially empty optical mode (vacuum state $|0\rangle_{\text{ph}}$), the joint state of the photonic state and a comb electron is $|\Psi_{\text{in}}\rangle = |\text{comb}_N^0\rangle_e \otimes |0\rangle_{\text{ph}}$. The interaction is described by the scattering matrix $S$ from Eq. (1) that acts on the joint state and creates (RM 2.2):

$$|\Psi_{\text{out}}\rangle = S|\Psi_{\text{in}}\rangle = \frac{1}{N}\sum_{k=0}^{N-1} c_N^k|\text{cat}_N^k\rangle_{\text{ph}} \otimes |\text{comb}_N^{-k}\rangle_e, \qquad (2)$$

where $|\text{cat}_N^k\rangle_{\text{ph}} = \frac{1}{c_N^k}\sum_{m=0}^{N-1} e^{-i2\pi mk/N}|g_Q e^{2im\pi/N}\rangle_{\text{ph}}$ is the $k^{\text{th}}$ order of the N-component cat state [9] and $c_N^k$ is a normalization factor that captures the probability of post-selecting the $k^{\text{th}}$ cat.

After the interaction, we post-select the electron energy to have a certain value $k \cdot \hbar\omega$ (modulo $N \cdot \hbar\omega$), which heralds the emission of a cat state $|\text{cat}_N^k\rangle_{\text{ph}}$ (Fig. 2a-f). For example, for the case of $N = 2$ (i.e., energy spacing $2\hbar\omega$) and post-selection of even/odd electron energies ($k = 0/1$), the electron radiation takes the form of the even/odd Schrodinger cat state (Fig. 2c,d), i.e., a superposition of two coherent states with opposite signs ($|g_Q\rangle_{\text{ph}}$ and $|-g_Q\rangle_{\text{ph}}$). This process of post-selecting even or odd cat states is analogous to a conditional displacement on the photonic mode, where the comb electron plays the role of the conditioning qubit. For any $N$ value, the amplitudes of the cat state components are proportional to the coupling constant $g_Q$. The probability to post-select a N-component cat state $|\text{cat}_N^k\rangle_{\text{ph}}$ is given by (RM 2.3)

$$P_N^k = \frac{1}{N^2}|c_N^k|^2 = \frac{1}{N^2}\left\|\sum_{m=0}^{N-1} e^{-i2\pi\frac{km}{N}}|e^{i2\pi\frac{m}{N}}g_Q\rangle_{\text{ph}}\right\|^2. \qquad (3)$$

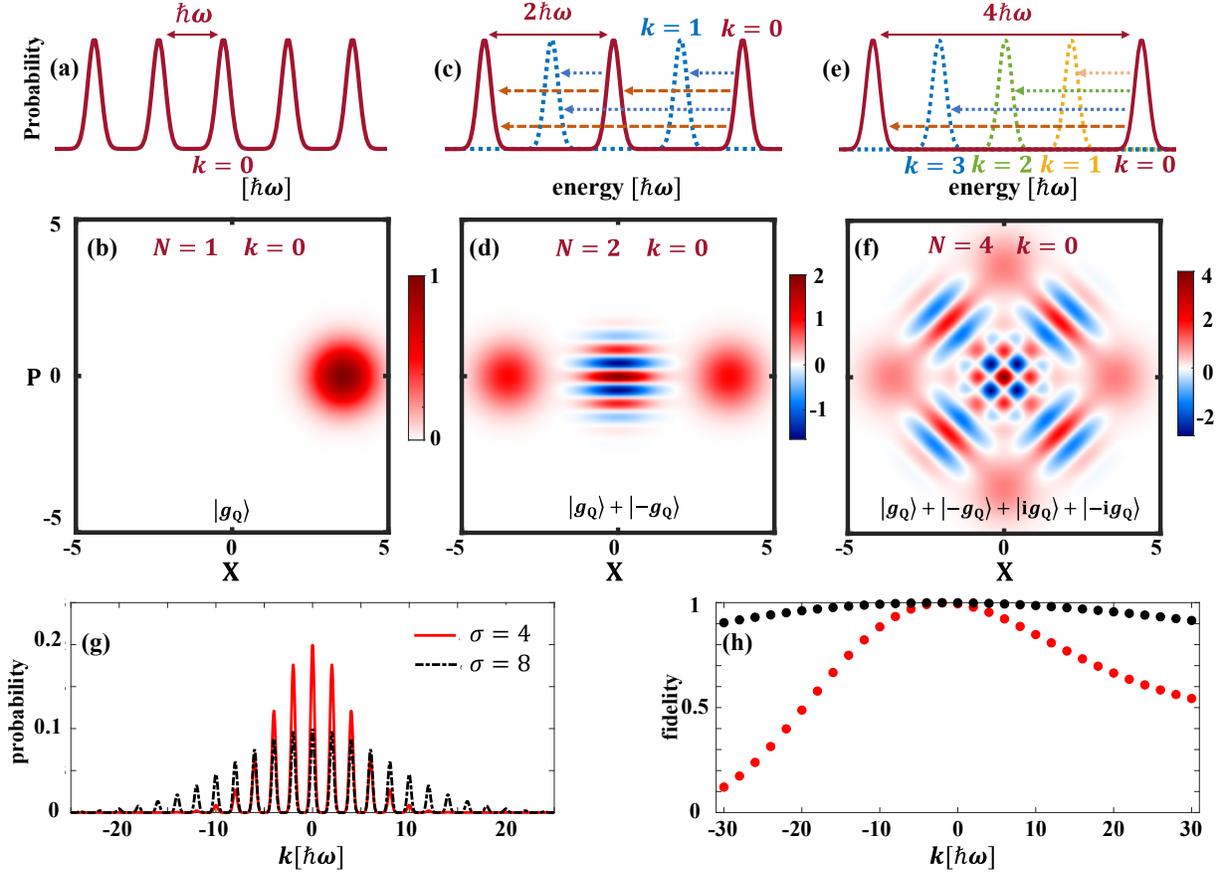

**Fig. 2 | Characterization of the $N$-component cat states emitted from a free electron. (a)** When a comb electron with an energy spacing of $\hbar\omega$ emits photons of energy $\hbar\omega$, **(b)** the Wigner function of the photon takes the form of a coherent state. **(c)** When a comb electron with an energy spacing of $2\hbar\omega$ is post-selected for even/odd energies after emitting photons of $\hbar\omega$, **(d)** the photonic Wigner function takes the form of even/odd cat states. **(e)** When a comb electron with an energy spacing of $4\hbar\omega$ is post-selected after emitting photons of $\hbar\omega$, **(f)** the photonic Wigner function takes the form of the different four-components cat states. **(g)** Energy spectra of Gaussian electron combs (energy spacing of $2\hbar\omega$) with a standard deviation of $\sigma = 4$ (red) and $\sigma = 8$ (black) in units of photon energy ($g_Q = \sqrt{\pi/2}$). Such states can be created with high fidelity by three PINEM-type interactions with classical laser light, as described in [69]. **(h)** The fidelity of the post-selected even cat states after interaction with the Gaussian comb electron. The average fidelity for $\sigma = 4$ is 0.97 and for $\sigma = 8$ is 0.99 (RM 5.2). ($g_Q = 4$ for all panels in this figure. Additional examples with lower $g_Q$ values are provided in the second panels of Figs. 3a, 4a, 4c – showing the creation of various kitten states).

### Creation of GKP states

To create a photonic state in a superposition of many coherent states, we consider multiple comb electrons with $2\hbar\omega$ spacing ($|\text{comb}_2^0\rangle_e$) interacting consequently with an optical mode. For commuting interactions, the electrons can arrive simultaneously as in a multi-electron pulse, under the condition of negligible electron-electron repulsion. For now, the optical mode is initiated with a vacuum state before the electron interactions. i.e., the electrons create the desired GKP states through a form of spontaneous emission, rather than a stimulated process.

Consider an interaction with $m + n$ comb electrons, with $n$ of them measured to have an odd energy and $m$ of them measured to have an even energy. For a general initial photonic state $|\psi_i\rangle_{\text{ph}}$, the final photonic state after the interaction is (RM 3.1):

$$|\psi_f\rangle_{\text{ph}} \propto \left(D_{g_Q} + D_{-g_Q}\right)^m \left(D_{g_Q} - D_{-g_Q}\right)^n |\psi_i\rangle_{\text{ph}}$$

$$\propto \sum_{n_1=0}^{m} \sum_{n_2=0}^{n} \binom{m}{n_1}\binom{n}{n_2}(-1)^{n_2} D_{g_Q(m+n-2n_1-2n_2)} |\psi_i\rangle_{\text{ph}}, \qquad (4)$$

where $D_{g_Q} = \exp(g_Q a^\dagger - g_Q^* a)$ is the displacement operator [8] and $\binom{m}{n_1} = \frac{m!}{n_1!(m-n_1)!}$ are the binomial coefficients (similarly to grid states proposed by [24]). Eq. (4) provides the possibility to generate different superpositions of coherent states including the squeezed vacuum state.

Superpositions of coherent states that form 2D grid states are possible if considering electron combs with energy spacing higher than $2\hbar\omega$, or by having two different interaction constants (RM 3.2). Among the different 2D grid states, the most attractive are the square and hexagonal GKP states [10,70,71]. These GKP states are desired since they enable fault-tolerant universal quantum computation with Gaussian operations [menicuci2013]. We propose several schemes for the creation of such states (Table 1).

The first scheme we present is for the creation of square GKP states. We choose $4m$ interactions of comb electrons ($|\text{comb}_2^0\rangle_e$ with coupling constant $g_{Q1} = i\sqrt{\pi/8}$. We post-select all of them to have even energies (Fig. 3a, first row). Then, we introduce $m$ additional interactions of electrons in the state $|\text{comb}_2^0\rangle_e$ with coupling constant $g_{Q2} = \sqrt{\pi/2}$, again post-selecting even energies (Fig. 3a, second row). Overall, the total number of electrons used in this scheme is $N_e = 5m$. In order to control the coupling constant's phase, one can change the phase of the laser used to shape the electron comb or change the region of the mode with which the electron interacts. For electrons with the same coupling constant phase, the order of interaction and order of post-selection do not matter because displacement with similar directions are commutative. This fact greatly simplifies the current scheme and the ones below

The resulting approximated GKP state is (RM 3.3):

$$|\text{GKP}'\rangle_{\text{ph}}^m \propto \sum_{n_1=0}^{m} \sum_{n_2=-0}^{4m} \binom{m}{n_1}\binom{4m}{n_2} D_{\sqrt{\frac{\pi}{2}}(2n_1-m)} D_{i\sqrt{\frac{\pi}{2}}(n_2-2m)} |0\rangle_{\text{ph}} \quad (5)$$

For an even/odd $m$, this state approximates the ideal GKP of a logical zero/one state $|0\rangle_{\text{ph}}^{\text{GKP}}$ / $|1\rangle_{\text{ph}}^{\text{GKP}}$. We recall that the ideal GKP states can be written as [70,71]:

$$|\mu\rangle_{\text{ph}}^{\text{GKP}} \propto \sum_{\vec{n}\in\mathbb{Z}} D_{\sqrt{\frac{\pi}{2}}(2n_1+\mu)} D_{i\sqrt{\frac{\pi}{2}}n_2} |0\rangle_{\text{ph}}, \quad (6)$$

where $\mu = 0/1$ defines the logical GKP qubits $|0\rangle_{\text{ph}}^{\text{GKP}}$ / $|1\rangle_{\text{ph}}^{\text{GKP}}$ respectively. When the number of electrons $m$ approaches infinity, the approximated state (Eq. (5)) approaches the ideal GKP (Eq. (6)).

**Fig. 3 | A scheme for creation of the GKP state: squeezing and post-selection probability.**
**(a)** The evolution of the Wigner function of the photonic state after each electron interaction and post-selection. The first interactions all have the same coupling constant $g_{Q1} = i\sqrt{\pi/8}$, together squeezing the vacuum state. We then shift the phase of the interaction by $\pi/2$, so the later interactions all have a coupling constant $g_{Q2} = \sqrt{\pi/2}$, transforming the squeezed-vacuum state into a GKP state. **(b)** Creation of GKP state directly from an initial squeezed vacuum excitation, i.e., "seeding" the electron-photon interaction with a squeezed vacuum in the optical mode. The GKP state is alternating between the approximated $|0\rangle_{\text{ph}}^{\text{GKP}}$ and $|1\rangle_{\text{ph}}^{\text{GKP}}$, showing that each interaction resembles the X gate for the GKP states. **(c)** The coefficients of the photonic state at every step of the process are described analytically using a Pascal triangle. This description simplifies the calculation of the post-selection probabilities in Eq. (6). (RM 4) **(d,e)** The squeezing parameter and post-selection probability of the final GKP state as a function of the number of electron interactions: comparing photonic initial conditions of vacuum (a) and squeezed vacuum (b).

To calculate the squeezing of the approximated state, we rewrite Eq. (5) in the $x$-quadrature representation (more details including the $p$-quadrature are found in RM 3.3):

$$\text{GKP}'(x) \propto \sum_{n_1=0}^{m} \binom{m}{n_1} \cos^{4m}\left(\sqrt{\pi}(x + m\sqrt{\pi})/2\right) e^{-\frac{1}{2}\left(x - \sqrt{\pi}(2n_1 - m)\right)^2}. \tag{7}$$

Eq. (7) describes a series of peaks with a distance of $2\sqrt{\pi}$, shifted by $0 \backslash \sqrt{\pi}$ for even\odd $m$. We note that the $\cos^{4m}$ term closely approximates a comb of Gaussian peaks (instead of delta-functions). The squeezing parameter is given by the variance of the peaks (of the corresponding probability distribution), which scales like $\Delta_x^2 \cong 1/(1 + \pi m)$. The corresponding squeezing parameter is defined as $S_{\text{dB}} = 10 \log_{10} 1/\Delta_x^2 = 10 \log_{10}(1 + \pi m)$, which thus grows logarithmically in the number of electrons. We choose $m$ interactions for $g_{Q2}$ and $4m$ for $g_{Q1}$ such that the squeezing is similar in the $x$ and $p$ representations. This way, Figs. 3d,e can present a single squeezing parameter by showing a datapoint every five electron interactions. The ideal GKP $|\mu\rangle_{\text{ph}}^{\text{GKP}}$ is obtained at the limit of $m \to \infty$. Substituting $m = 3$ shows that $N_e = 15$ electrons are required to achieve ~10 dB squeezing (Fig. 3d), which is the estimated squeezing level for fault-tolerant quantum computing (reaching the quantum error correction threshold) using CVs [72,73].

The post-selection probability to obtain the state $|\text{GKP}'\rangle_{\text{ph}}^m$ is (illustrated in Fig. 3e and detailed in RM 4.3):

$$P_{|\text{GKP}'\rangle_{\text{ph}}^m} = \frac{\left\|\left(D_{\sqrt{\pi/2}} + D_{-\sqrt{\pi/2}}\right)^m \left(D_{i\sqrt{\pi/8}} + D_{-i\sqrt{\pi/8}}\right)^{4m} |0\rangle_{\text{ph}}\right\|^2}{4^{5m}}. \tag{8}$$

The post-selection probability to produce the GKP of 10 dB squeezing ($m = 3$) according to Eq. (8) is ~10% (Fig. 3e). As expected, the probability in Eq. (8) decreases for larger $m$, i.e., for a larger number of electrons $N_e$. However, the probabilities decay rather slowly with $N_e$, like $\sim 5/(N_e \pi)$ (RM 4.3), which leaves us with relatively high success rates. This fact may seem somewhat surprising when recalling that the success probability of post-selecting the first electron is close to 50%, and that multiple post-selections often scale exponentially in this probability. An exponential scaling would have caused the entire scheme to be impractical, and thus it is highly encouraging to instead find a power-law scaling in the number of electrons.

To increase the success probability of creating a GKP state further, one can stimulate the interaction with a squeezed vacuum state (Fig. 3b). We consider a squeezed vacuum state in the initial photonic mode before the interaction $|\psi_i\rangle_{\text{ph}} = S(\xi)|0\rangle_{\text{ph}}$, with $S(\xi)$ being the squeezing operator $\exp\left(\frac{1}{2}\xi^*\hat{a}^2 - \frac{1}{2}\xi\hat{a}^{\dagger 2}\right)$, and $\xi = re^{i\theta}$ being the squeezing parameter [8]. For "seeding" a squeezed vacuum into the optical mode, one can use mature techniques like spontaneous parametric down-conversion or spontaneous four-wave mixing. After the interaction of $N_e$ electrons with the squeezed vacuum state (where $g_Q = \sqrt{\pi/2}$, $\theta = 0$), the resulting photonic state becomes (RM 3.4):

$$|\text{GKP}''\rangle_{\text{ph}}^{N_e} \propto \sum_{n=0}^{N_e} \binom{N_e}{n_1} D_{\sqrt{\frac{\pi}{2}}(2n_1 - N_e)} S(\xi)|0\rangle_{\text{ph}}, \tag{9}$$

which is an approximation of GKP states [24]. Writing Eq. (9) in the $p$-representation:

$$\text{GKP}''(p) \propto \exp\left(-\frac{p^2}{2}e^{-2r}\right)\left(1 + e^{-2i\sqrt{\pi}p}\right)^{N_e}. \tag{10}$$

A comparison with the $x$-representation is discussed in RM 3.4. The probability of post-selecting all electrons with even energies is (RM 4.4):

$$P_{|\text{GKP}''\rangle_{\text{ph}}^{N_e}} = \frac{1}{4^{N_e}}\sum_{n=0}^{2N_e}\binom{2N_e}{n}e^{-\pi(N_e-n)^2|\cosh r + \sinh r|^2}. \tag{11}$$

The probability here decays more slowly, like $1/\sqrt{N_e\pi}$, because there is one axis rather than two for the displacement interactions.

We calculate the squeezing parameter of the optical GKP state as a function of the number of electrons $N_e$ and find $S_{\text{dB}} = 10\log_{10}(e^{-2r} + N_e\pi)$ [RM 3.4]. To achieve 9.8dB squeezing for the GKP state (Fig. 3d), we need just three electrons. The post-selection probability to produce this state according to Eq. (11) is 31.25% (Fig. 3e).

The reason for the relatively high success probabilities is that the electron-photon scattering matrix $S$ of Eq. (1) causes the quantum state to gradually converge into the ideal GKP states. The closer the photonic state reaches, the better the success probability becomes. Destructive interference in the electron wavefunction reduces the probability of the electron acquiring odd energies after the interaction, and thus increases the post-selection probability. The sequential application of the electron interaction and post-selection (with the coupling constant $g_{Q2} = \sqrt{\pi/2}$) causes a convergence into the GKP states, shifting between the $|0\rangle_{\text{ph}}^{\text{GKP}}$ and the $|1\rangle_{\text{ph}}^{\text{GKP}}$ GKP states for even and odd number of interactions respectively. Using the terminology of quantum error correction, two consequent interactions are a stabilizer for the GKP state [10] (RM 7). We find a similar convergence for the other GKP states for different interaction parameters as listed in Table 1. In all these cases, the interactions with comb electrons define stabilizers for the corresponding GKP states, making such fundamental interactions precisely suited for the creation of GKP states.

| | Initial state | Interaction description | $g_{Q,\max}$ | $P_{10\text{dB}}$ [%] | $N_e$ | Post-selection | Final state |
|---|---|---|---|---|---|---|---|
| 1 | $|0\rangle_{\text{ph}}$ | $\sum_{n_1=0}^{2m}\sum_{n_2=0}^{2m}\binom{2m}{n_1}\binom{2m}{n_2} D_{i\sqrt{\frac{\pi}{2}}(n_1-m)} D_{\sqrt{\frac{\pi}{2}}(n_2-m)}$ | $\frac{1}{2}\sqrt{\frac{\pi}{2}}$ | 5 | 24 | E E | $|0\rangle_{\text{ph}}^{\text{GKP}}$ |
| 2 | $|0\rangle_{\text{ph}}$ | $\sum_{n_1=0}^{m}\sum_{n_2=0}^{4m}\binom{m}{n_1}\binom{4m}{n_2} D_{\sqrt{\frac{\pi}{2}}(2n_1-m)} D_{i\sqrt{\frac{\pi}{2}}(n_2-2m)}$ | $\sqrt{\frac{\pi}{2}}$ | 9.7 | 15 | E E | $|0/1\rangle_{\text{ph}}^{\text{GKP}}$ |
| 3 | $|0\rangle_{\text{ph}}$ | $\sum_{n_1=0}^{m}\sum_{n_2=0}^{m}\binom{m}{n_1}\binom{m}{n_2} D_{\sqrt{\frac{\pi}{2}}(2n_1-m)} D_{i\sqrt{\frac{\pi}{2}}(2n_2-m)}$ | $\sqrt{\frac{\pi}{2}}$ | 11.1 | 6 | E E | $|H\rangle_{\text{ph}}^{\text{GKP}}$ |
| 4 | $|0\rangle_{\text{ph}}$ | $\sum_{\alpha=0}^{2m}\sum_{\beta=0}^{2m}\sum_{\gamma=0}^{4m}\binom{2m}{\alpha}\binom{2m}{\beta}\binom{4m}{\gamma}(-1)^\beta D_{\frac{1}{2}\sqrt{\frac{\pi}{2}}(\alpha+\beta-2m)} D_{\frac{i}{2}\sqrt{\frac{\pi}{2}}(\gamma-2m)}|0\rangle$ | $\frac{1}{4}\sqrt{\frac{\pi}{2}}$ | 0.4 | 96 | E O E | $|-\rangle_{\text{ph}}^{\text{GKP}}$ |
| 5 | $S(\xi)|0\rangle_{\text{ph}}$ | $\sum_{n_1=0}^{m}\binom{m}{n_1} D_{\sqrt{\frac{\pi}{2}}(2n_1-m)} S(r=1.1513, \theta=0)|0\rangle$ | $\sqrt{\frac{\pi}{2}}$ | 31.3 | 3 | E | $|0/1\rangle_{\text{ph}}^{\text{GKP}}$ |
| 6 | $|0\rangle_{\text{ph}}$ | $\sum_{n_1=0}^{2m}\sum_{n_2=0}^{2m}\binom{2m}{n_1}\binom{2m}{n_2} D_{e^{2i\pi/3}\sqrt{\frac{\pi}{\sqrt{3}}}(n_1-m)} D_{\sqrt{\frac{\pi}{\sqrt{3}}}(n_2-m)}|0\rangle$ | $\frac{1}{2}\sqrt{\frac{\pi}{\sqrt{3}}}$ | 2.6 | 44 | E E | $|0\rangle_{\text{ph}}^{\text{GKP}}$ |
| 7 | $|0\rangle_{\text{ph}}$ | $\sum_{n_1=0}^{m}\sum_{n_2=0}^{m}\sum_{n_3=0}^{m}\binom{m}{n_1}\binom{m}{n_2}\binom{m}{n_3} D_{\sqrt{\frac{\pi}{\sqrt{3}}}(2n_1-m)} D_{e^{2i\pi/3}\sqrt{\frac{\pi}{\sqrt{3}}}(2n_2-m)} D_{e^{4i\pi/3}\sqrt{\frac{\pi}{\sqrt{3}}}(2n_3-m)}|0\rangle$ | $\sqrt{\frac{\pi}{\sqrt{3}}}$ | 9.5 | 6 | E E E | $|T\rangle_{\text{ph}}^{\text{GKP}}$ |
| 8 | $S(\xi)|0\rangle_{\text{ph}}$ | $\sum_{n_1=0}^{m}\binom{m}{n_1} D_{\sqrt{\frac{\pi}{\sqrt{3}}}(2n_1-m)} S\left(r=1.64, \theta=\frac{\pi}{6}\right)|0\rangle$ | $\sqrt{\frac{\pi}{\sqrt{3}}}$ | 27.3 | 4 | E | $|0/1\rangle_{\text{ph}}^{\text{GKP}}$ |

**Table 1 | Different protocols for the creation of grid coherent states using electron combs.** Rows 1-4 describe different protocols for the creation of approximated square GKP states, with different $g_Q$ values. Row 5 shows a protocol for the creation of approximated square GKP states when starting from squeezed vacuum rather than from vacuum. Rows 6-8 show similar protocols for the creation of approximated hexagonal GKP states. $g_{Q,max}$ refers to the highest coupling constant used as part of the protocol. $P_{10dB}$ is the probability to achieve GKP with ~10dB squeezing. $N_e$ is the number of electron interactions required for achieving this squeezing value. The post-selection column describes the sequence of post-selections necessary to create the state, where E/O stands for even/odd electron energies.

**Creation of hexagonal GKP and magic states**

Our approach enables the creation of additional types of GKP states such as the hexagonal GKP (Fig. 4b). Examples are summarized in Table 1, each requiring different coupling constants with different relative phases between the sets of interactions. The magic GKP states (Fig. 4a,c), first proposed by [74], enable universal quantum computation without requiring additional non-Gaussian elements [75]. We show how magic states can be created by using comb electrons. For example, for the square GKP magic state $|H\rangle_{ph}^{GKP}$ shown in Fig. 4a, we propose a scheme involving $N_e = 2m$ electrons (presented in the 3$^{rd}$ row of Table 1): having $m$ interactions with coupling $g_{Q1} = i\sqrt{\pi/2}$ post-selected for even energies, followed by additional $m$ interactions with coupling $g_{Q2} = \sqrt{\pi/2}$ post-selected for even energies. Similarly, the hexagonal GKP magic state $|T\rangle_{ph}^{GKP}$ shown in Fig. 4c can be created as shown in the 7$^{th}$ row of Table 1.

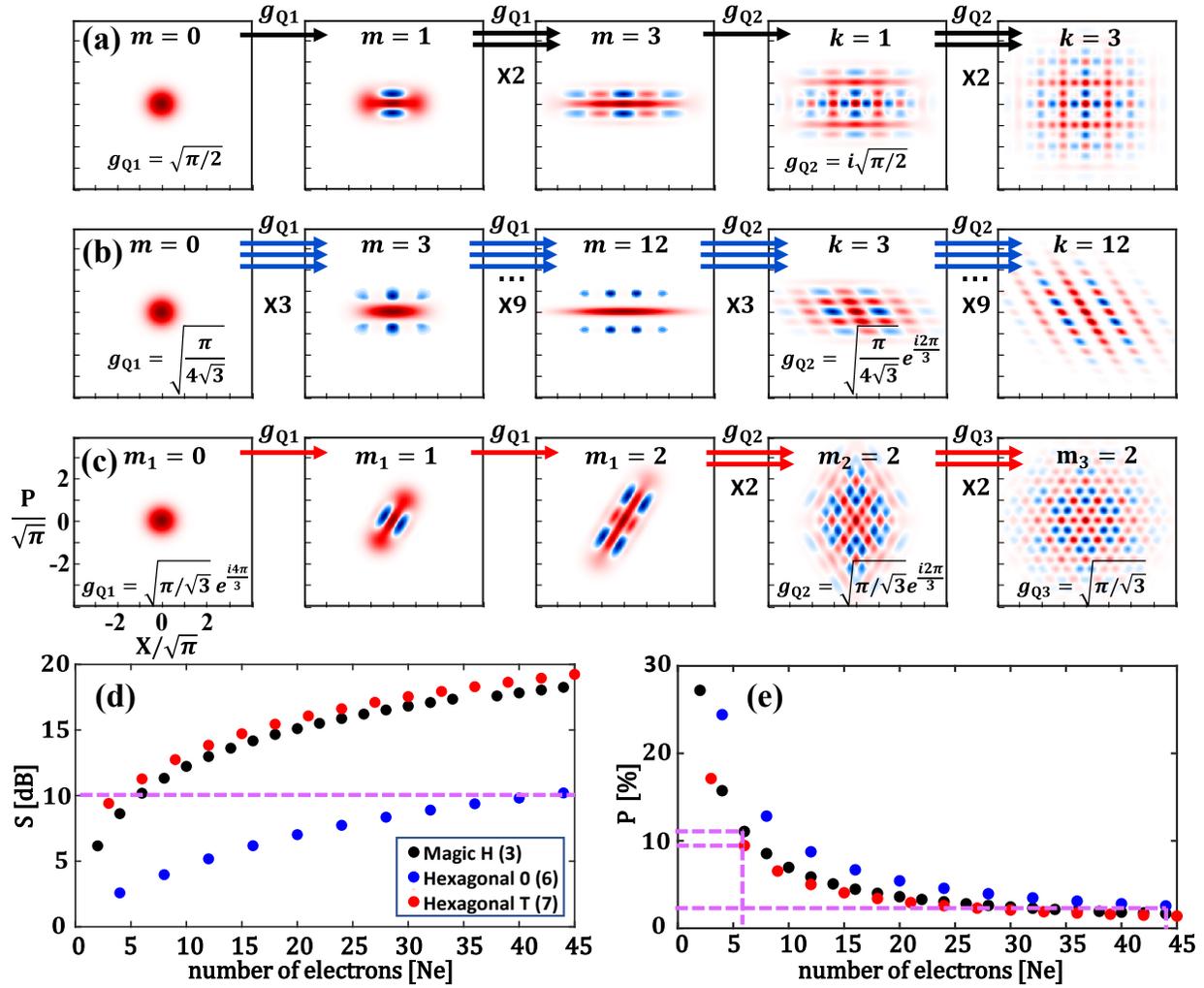

**Fig. 4 | Schemes for creation of different GKP states and their characterization. (a)** Creation of the magic GKP state $|H\rangle_{ph}^{GKP}$ (3$^{rd}$ row in Table 1). **(b)** Creation of the hexagonal GKP state $|0\rangle_L$ (6$^{th}$ row in Table 1). **(c)** Creation of the hexagonal magic state $|T\rangle_{ph}^{GKP}$ (7$^{th}$ row in Table 1). **(d,e)** The squeezing parameter and post-selection probability of the GKP states in (a,b,c) as a function of the number of electrons.

**Creation of entangled GKP states – toward cluster states**

It is insightful to recast the electron-photon interaction to the language of quantum gates. Specifically, the same comb electrons used above to create the GKP state enable implementing quantum gates such as the Pauli X, Y, and Z for the GKP states (RM 7). We now combine this approach with the ideas developed in [76] to induce entanglement between two photonic modes. A free-electron interaction with two photonic modes can entangle them by performing two-qubit gates (e.g., CNOT), creating a GKP Bell state. To see that, we consider an electron that interacts with two photonic modes, e.g., by placing two cavities along the electron trajectory. The combined interaction is then described by two scattering matrices $S_1, S_2$, each related to the interaction with a different photonic mode.

As an example, consider the following initial state:

$$|\psi_{\text{initial}}\rangle = |0\rangle_{\text{ph1}}^{\text{GKP}} |0\rangle_{\text{ph2}}^{\text{GKP}} |\text{comb}_4^0\rangle_e, \tag{14}$$

which corresponds to an electron comb with a spacing of $4\hbar\omega$ and two photonic modes ($|0\rangle_{\text{ph1}}^{\text{GKP}}, |0\rangle_{\text{ph2}}^{\text{GKP}}$) in a GKP state. Following the interaction of this electron comb with both modes, we post-select the electron energy. The result is (see RM 6):

$$\begin{cases} |+\rangle_{\text{ph1}}^{\text{GKP}} |+\rangle_{\text{ph2}}^{\text{GKP}} + |-\rangle_{\text{ph1}}^{\text{GKP}} |-\rangle_{\text{ph2}}^{\text{GKP}}, & \text{for post-selecting } |\text{comb}_4^0\rangle_e \\ |+\rangle_{\text{ph1}}^{\text{GKP}} |-\rangle_{\text{ph2}}^{\text{GKP}} + |-\rangle_{\text{ph1}}^{\text{GKP}} |+\rangle_{\text{ph2}}^{\text{GKP}}, & \text{for post-selecting } |\text{comb}_4^2\rangle_e \end{cases} \tag{15}$$

Both options are GKP Bell states. Post-selecting the electron will generate one of these Bell states, according to the measured electron energy. The electron state $|\text{comb}_4^0\rangle_e$ acts like a conditional rotation gate for the GKP square state (similarly, for the hexagonal-GKP state, a $|\text{comb}_6^0\rangle_e$ can be used). Looking forward, this scheme can be generalized to entangle a larger number of states and even to create GKP cluster states.

**Fidelity estimation**

We now turn to estimate the fidelity of the cat and GKP states. Different considerations can lower the fidelity in practical settings, including detectors efficiencies, deviation from ideal comb, variance in the constant coupling $g_Q$, the bandwidth of the optical mode (or multimode), dispersion of either the electron or the photonic modes, aberrations for the temporal and transversal electron beam focusing, and electron-electron repulsion. That said, we can give a strong indication of the robustness of the proposed approach. We consider a standard deviation $\Delta g_Q$ in the value of $g_Q$ (the variation is taken to be in the amplitude for this example). Such deviations in PINEM experiments result from transverse non-uniformities in the field $E_z(r_\perp, z)$, or changes of the interaction length. We find that the fidelity for the case of a two-component cat state goes like $\propto 1 - \Delta g_Q^2$, with $\Delta g_Q < 0.25$ (RM 5.5 and Fig. RM1b).

The fidelity is also limited by the quality of the comb electron. Any Gaussian comb has two characterizing features: its envelope width and the energy width of the individual peaks (coherent and incoherent broadening). The width of each energy peak can create an error in the detection due to some overlap between adjacent peaks. For high fidelity, the ratio between the energy of the photon and the standard deviation of each peak should be above three standard deviations for error rates below 1%. For the telecom range, the photon energy is 0.8ev, this means that the standard deviation of the electron energy width of each peak should be ~ 0.13 eV (~0.3eV FWHM), which is achievable in TEM and even UTEM [77]).

Another consideration is the finite energy width of realistic electron combs that approximate the infinite width of the ideal comb. A wider electron comb increases the fidelity (RM 5.3, Fig. 2h), but only up to a certain propagation distance, because a wider comb experiences stronger dispersion that distorts the phases of the comb peaks, limiting the fidelity. Strong coupling ($g_Q > 0.1$), as necessary for GKP generation, requires long phase-matched interactions [62,55,78], on the order of a hundred microns (for 200 keV electron combs and photons of 1550 nm). Since the phase-distortion by dispersion grows linearly with the distance (RM 5.6), there exists a different optimal electron energy width for each interaction distance and each required fidelity. i.e., too wide a comb will in fact smear the phase due to electron dispersion, resulting in lower fidelities (RM 5.6, Fig. RM 1c). These considerations create an inherent trade-off between the fidelity and coupling strength $g_Q$. Our results show that despite this trade-off, there is a wide range of parameters for which we can create the photonic states necessary for CV fault-tolerant quantum computing.

To provide concrete fidelity estimates for GKP states created by approximated electron combs, we consider a Gaussian envelope for the electron energy spectrum as in Ref. [69]). The Gaussian envelope is preferable since it maximizes the first moments while minimizing higher-order moments of the electron energy spectrum. We estimate that creating an approximate GKP state (Eq. (5)) with 98% fidelity requires a standard deviation of 30 peaks (RM 5.5). As an example, we consider such an electron with mean kinetic energy of 200 keV. We choose the distance between the electron energy peaks to match the energy of a photon at 775 nm, so the emission is into a GKP state at 1550 nm. The fidelity of the GKP state created by such a comb can remain above 97% for an interaction distance of up to 160 μm. Such a distance is sufficient to create the strong coupling strength $g_Q$ as was predicted in [62] and shown in [78]. These papers and the others discussed below show that each of the necessary components toward the realization of our proposal has been demonstrated in a separate experiment in recent years.

Taken together, these advances help us envision a roadmap toward the full experimental demonstration of free-electron generation of optical GKP states.

## Discussion

**Roadmap toward an experimental realization**

The realization of free-electron-driven optical GKP states requires addressing important challenges in each of the three stages of the process (Fig. 1a-c). In the preparation stage (Fig. 1a), multiple harmonics [68] or multiple interactions [69] are necessary to shape the electrons into high-quality combs. Recently, strong electron shaping with a continuous-wave laser was demonstrated [57,58], instead of the short laser pulses used in previous experiments of this kind [38,39,40,55,51,41,53]. This mode of operation allows for coherent electron shaping (i.e., temporal modulation) with less complicated synchronization for the interactions between the electron and the shaping light. However, for the continuous-wave interaction to be efficient enough (the combs we consider in Fig RM 5.6 have coupling $g = \sqrt{N} g_Q \sim 50$, with $N$ being the mean number of photons), Refs. [57,58] utilized grazing angle conditions for phase-matched or quasi-phase-matched interactions. Such grazing angle conditions require strong electron lenses to create a small electron beam diameter together with small convergence angles. These conditions can be met in specialized TEM with multiple objective lenses.

The light emission stage (Fig. 1b) requires strong coupling ($g_Q > 0.5$) between the electrons and the optical mode. To our knowledge, the highest $g_Q$ reported so far was close to unity [78], which is at the scale of the values needed for GKP generation. Such a $g_Q$ value was demonstrated using a structure supporting hybrid modes of surface plasmon polaritons and photons in a waveguide [78]. The disadvantage of this scheme is the lossy nature of such polaritons. Theoretical proposals for similar coupling efficiencies with smaller losses are based on electron interaction with micro-cavities [62], or with photonic crystal flat-bands [79]. The value of $g_Q$ can be further increased, even much above unity, using a longer phase-matched interaction and a highly confined optical mode [62]. Interaction lengths of up to 500 μm have been demonstrated [55], and we estimate that order of 150 μm is required for a large enough $g_Q$. Moreover, the optical mode must have a substantial part of its energy in a vacuum and have a longitudinal polarization to ensure efficient evanescent coupling to the electron (as shown in [78]). These interaction conditions also depict the timescale over which the emission occurs, which in the case of a 160 μm structure and 200 keV electrons is ~800 fs. This timescale also corresponds to the bandwidth of the emitted radiation, which is about ~10 nm for emission at 1550 nm. The photonic losses should be negligible over the interaction length and time scale, as is indeed the case in state-of-the-art dielectric waveguides and microcavities.

An important consideration for the light emission stage arises from our assumption that the electron interacts with only one optical mode. i.e., the electron coupling strength ($g_Q$) with a specific mode is much larger than with other modes. Satisfying this assumption necessitates a carefully designed structure, which might be achieved by exploiting methods of nanophotonic design. We can also consider enhancing the interaction with a single desired mode by pumping it with a weak coherent state, or a weakly displaced squeezed vacuum. The alternative solution is creating the GKP states such that each photon is a superposition of multiple spectral modes. In such scenario, the coherent width of each electron energy peak needs to be wider than the spectral width of the photon emission. The coherent energy width of electrons emitted from typical cathodes is on the order of 0.3 eV [80,81]. The phase-matching condition imposes a minimal interaction length of ~30 μm to reduce the spectral width between the emitted modes to below 0.3 eV.

<u>The electron post-selection stage (Fig. 1c)</u> necessitates coincidence measurements and direct detection of individual electrons. Such measurements are already possible due to advances in electron counting direct detectors. Recent experiments reported coincidence of electron energy loss with X-ray emission [82] and with optical photon emission [32,83] using single-photon detectors. One advantage of free-electron generation of GKP states arises from the developments in fast electron counting detectors (direct detection schemes). Since free electrons are energetic particles, it is in principle easier to detect multiple electrons than achieve a similar detection with photons. This enables measuring simultaneous/rapid multi-electron interactions as used in our schemes above.

## Summary and outlook

In summary, this paper demonstrates how shaping and post-selecting free electrons can generate *N*-component cat states and grid states that are desired for CV quantum information processing. Furthermore, we show how multiple electrons can create GKP states in an approach that resembles breeding schemes [23-25]. Recent experimental achievements show the feasibility of our proposal, for example, the necessary pre-shaping of free electrons into wide coherent energy combs was demonstrated in several recent experiments [55,58,57]. Controlling the comb spacing can be achieved using a PINEM-type interaction, driven by harmonics of the fundamental laser frequency [40,68] or by multiple points of interaction [69]. In future works, we envision combining transverse and longitudinal shaping to scale this scheme to large GKP cluster states.

It is interesting to search for other opportunities to apply the schemes we developed here in places where a similar scattering matrix arises. Any physical mechanism that is described by such a scattering matrix may now be facilitated for the creation of quantum states of light. For example, a single-photon optical frequency comb going through a $\chi^{(2)}$ medium can generate photonic squeezed states at a frequency corresponding to half the difference between the spectral peaks of the comb (typically MHz to THz, depending on the frequency comb). The emitted photonic squeezed state is then heralded by a spectrometry measurement of the frequency comb photon. If a large-enough $\chi^{(2)}$ could be achieved (corresponding to a large $g_Q$), our approach would enable to directly create photonic GKP states using single-photon frequency combs.

Alternative routes of using the approach in this work are exploiting existing experiments that already create bunched electron pulses and beams. For example, above-threshold ionization and free-electron lasers create bunched electrons beams that can often contain multiple electrons per bunch. We propose launching bunched electrons with temporal modulation of $\pi/\hbar\omega$ the state $|\text{comb}_2^0\rangle_e$ (i.e., energy spacing $2\hbar\omega$, or ) into an undulator with double the period ($2\pi/\hbar\omega$), to trigger electron radiation at frequency $\hbar\omega$. In such a scenario, our work predicts that the resulting undulator radiation will be squeezed and heralded by measuring the electrons' energy. This approach can create squeezed X-ray states using highly relativistic electrons. Moreover, if a large effective $g_Q$ can be achieved in undulators (e.g., using highly charged ions [84]), then such undulator radiation would take the form of cat states and GKP states. This way, we envision future free-electron lasers that create cat states in spectral regions such as THz and X-rays, where no other methods currently exist for the creation of such quantum states.

Taking a wider perspective, the comb electrons can themselves encode quantum information, as shown by the recent proposal [85] and observation [56] of free-electron qubit schemes. We note that optical GKP states are analogous to electron combs with a spacing of $2\hbar\omega$, where the logical $|0\rangle_e$ ($|1\rangle_e$) state is the even (odd) electron energy comb. This analogy has certain similarities to the idea of encoding CV quantum information on a single-photon frequency comb [86]. Using this description, each comb electron can be thought of as an ancilla qubit, which through its interaction with the optical mode performs a conditional displacement on the photonic state. The conditional displacement together with regular displacements enables quantum error correction of GKP qubits as in Ref. [15]. Single-qubit universal control of such electrons was shown in Ref. [85]. Using such operations, our work provides a novel mechanism for quantum error-correction schemes based on free-electron interactions. Such interactions provide the necessary components toward a vision of free-electron-assisted fault-tolerant photonic quantum computation in the optical region.

**ACKNOWLEDGMENTS**

We thank Asaf Diringer and Aviv Karnieli for the valuable conversations and insight. The research was supported by the European Research Council (ERC Starting Grant 851780-NanoEP), the European Union Horizon 2020 Research and Innovation Program (under grant agreement No. 964591 SMART-electron). R.D. would like to acknowledge the support of the Council for Higher Education Support Program for Outstanding Ph.D. Candidates in Quantum Science and Technology in Research Universities. G.B. would like to acknowledge the support of the Technion Excellence Program for undergraduate students. I.K. and A.G acknowledge the support of the Azrieli Fellowship. This research was supported by the Technion Helen Diller Quantum Center.